\documentclass[a4paper,english,12pt]{article}
\usepackage{babel}
\usepackage{graphicx}
\usepackage{amsmath}
\usepackage{amsfonts}
\usepackage{amssymb}
\usepackage{dsfont}
\usepackage[hang]{caption2}
\usepackage{geometry}
\usepackage{hyperref}

\begin{document}
\begin{center}
\huge\textbf{{Inhomogeneous Quantum Mixmaster: from Classical toward Quantum Mechanics}}
\end{center}
\markboth{R. Benini, G. Montani}
{}

\begin{center}
\textbf{Riccardo Benini}\\
\footnotesize{Dipartimento di Fisica - Universit\`a di Bologna and INFN - Sezione di Bologna,
via Irnerio 46, 40126 Bologna, Italy\\
ICRA---International Center for Relativistic Astrophysics  
c/o Dipartimento di Fisica (G9) Universit\`a di Roma ``La Sapienza'',
Piazza A.Moro 5 00185 Roma, Italy}\\
\small{\textit{riccardo.benini@icra.it}}
\\

\textbf{Giovanni Montani}\\
\footnotesize{Dipartimento di Fisica Universit\`a di Roma ``La Sapienza''\\
ICRA---International Center for Relativistic Astrophysics  
c/o Dipartimento di Fisica (G9) Universit\`a di Roma ``La Sapienza'',
Piazza A.Moro 5 00185 Rome, Italy\\
ENEA C.R. Frascati (U.T.S. Fusione), Via Enrico Fermi 45, 00044 Frascati, Roma, Italy}\\
\small{\textit{montani@icra.it}}
\end{center}

\begin{abstract}

Starting from the Hamiltonian formulation for the inhomogeneous Mixmaster dynamics, we approach its quantum features through the link of the quasi-classical limit. We fix the proper operator-ordering which ensures that the WKB continuity equation overlaps the Liouville theorem as restricted to the configuration space.
We describe the full quantum dynamics of the model  in some details, providing a characterization of the (discrete) spectrum with analytic expressions for the limit of high occupation number. One of the main achievements of our analysis relies on the description of the ground state morphology, showing how it is characterized by a non-vanishing zero-point energy associated to the Universe anisotropy degrees of freedom

\end{abstract}


\section{Introduction}

Einstein's equations, despite their non-linearity,  admit a generic solution asymptotically to the cosmological singularity having a piece-wise analytic expression. This dynamical regime was derived by V. A. Belinski, I. M. Khalatnikov and E. M. Lifshitz (BKL)\cite{BKL70,BKL82} and it has an oscillatory-like behavior which extends point by point in space the homogeneous Mixmaster dynamics of the Bianchi type IX model \cite{M69}. When a generic initial condition is evolved toward the initial singularity, a space-time foam\cite{K93,M95} appears: this  is a direct consequence of the dynamical decoupling which characterizes space points close enough to the singularity and which can be appropriately described by the Mixmaster point like measure \cite{K93,KM97p} (a wide literature exists about the chaotic properties of the homogeneous Mixmaster model; see, for instance, \cite{BKL70,CB83,KM97J,CL97,IM01} and the references therein).\\
However, this classical description  is in conflict with the requirement of a quantum behavior of the Universe through the Planck era; in fact there are reliable indications \cite{KM97J} that the Mixmaster dynamics overlaps the quantum Universe evolution, requiring an appropriate analysis of the transition between these two different regimes.\\
Indeed the dynamics of the very early Universe corresponds to a very peculiar situation with respect to the link existing between the classical and quantum regimes because the expansion of the Universe is the crucial phenomenon which maps into each other these two stages of the evolution. As shown in \cite{KM97J} the appearance of a classical background takes place essentially at the end of the Mixmaster phase when the anisotropy degrees of freedom can be treated as small perturbations; this result indicates that the oscillatory regime takes place almost during the Planck era and therefore it is a problem of quantum dynamics. However the end of the Mixmaster (and in principle the {\it quantum to classical} transition phase) is fixed by the initial conditions on the system and in particular it takes place when the cosmological horizon reaches the inhomogeneity scale of the model; therefore the question of an appropriate treatment for the semiclassical behavior arises when the inhomogeneity scale is so larger than the Planck scale that the horizon can approach it only in the classical limit.
Some interesting features of the quantum Mixmaster have been developed by \cite{M69q, FU86, BE89, GRAHAM95} and \cite{K92, IM05} for the homogeneous and the inhomogeneous cases, respectively.\\ 
Here we refine this analysis both connecting some properties of the quantum behavior to the ensemble representation of the model, both describing the precise effect of the boundary conditions on the structure of the quantum states.\\
We start by an Arnowitt-Deser-Misner (ADM)\cite{ADM62} representation of the system dynamics which allows us to disregard the contributions of the spatial gradients relative 
to the configurational variables
, thus reducing the dynamics to a number of $\infty^3$ independent point-like Mixmaster model. Such  representation can take place independently of the gauge choice\cite{BM04}, nevertheless here we require  one of the Misner-Chitr\'e like variables to play the role of time for the system. This way, we can describe the whole evolution in terms of a triangular billiard on the Poincar\'e upper half-plane where the point Universe randomizes.\\
Then we apply the Liouville theorem restricted to the configuration space using the Hamilton-Jacobi solution of the model, and and we require the continuity equation to match the WKB limit. Such a requirement leads us to determine the proper operator-ordering associated to the ADM operator in a unique way. We emphasize that the existence of an energy-like constant of motion not only provides a microcanonical measure for the statistical dynamics but also induces a quantum dynamics completely described by a time-independent Schr\"odinger equation on the Poincar\'e plane.
However, the effective Hamiltonian associated to such an eigenvalue problem is non-local, characterized by the presence of a square root function; in agreement with the  analysis developed by \cite{Puzio}, we make the well-grounded hypothesis that the eigenfunctions  form is independent of the presence of the square root, since its removal implies the square of the eigenvalues only.\\
A crucial step relies on recognizing how our squared equation can be restated as the Laplace-Beltrami eigenvalue problem.
Hence a characterization of the "energy" spectrum comes out from imposing the triangular boundary conditions (the semicircle is replaced by a straight line preserving the measure of the domain).\\
We outline that the spectrum is discrete and that admits a point-zero energy which implies an intrinsic anisotropy of the quantum Mixmaster Universe. We numerically investigate the first "energy" levels and finally provide an analytic expressions for high occupation numbers.\\
The structure of the presentation is the following: Section II is devoted to fix the classical Hamiltonian formulation of the system dynamics, based on the use of Misner-Chitr\'e-like variables which allow the ADM reduction of the degrees of freedom.
In Section III we fix the solution of the Hamilton-Jacobi equation within the domain allowed by the vanishing of the potential.
Such a solution is at the ground level of the analysis presented in Section IV; in fact here we restrict the Liouville theorem to the configuration space, eliminating the momentum dependence by virtue of an integration along the Hamilton-Jacobi dynamics.
In Section V we describe the Schr\"odinger equation for the Mixmaster model and construct the WKB limit; by the comparison of this semiclassical behavior with the equation induced by the Liouville theorem, we get a unique choice for the operator-ordering of the super-Hamiltonian kinetic term.\\
Finally, in Section VI we provide a detailed description of the Mixmaster spectrum and eigenfunctions; particular attention is devoted to the ground state in view of the idea that it is the expected Universe configuration during the Planck era.
Section VII is devoted to concluding remarks on the physical issues of our discussion, and particular attention is devoted to fix the full inhomogeneous  quantum picture.\\
Over the whole paper we adopt units such that $c=16\pi G=1$.

\section{Hamiltonian Formulation}

\begin{figure} 
\begin{center} 
\includegraphics[width=8.0cm]{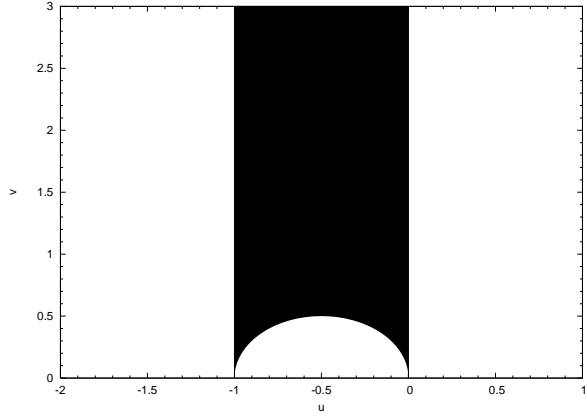} 
\caption[prrr]{The domain $\Gamma_Q(u,v)$ in the Poincar\'e upper half-plane where the dynamics of the space-point Universe is asymptotically restricted by the potential.\protect\\ The measure is finite and equal to $\pi$. \label{pot1.eps}} 
\end{center}
\end{figure}

In this section we introduce the Hamiltonian formalism for a generic inhomogeneous model in  vacuum in terms of the Misner-Chitr\'e-like variables \cite{M69,C72}.\\
In the ADM approach, the line element associated to such a model can be written in the form:
\begin{equation}
	ds^2=N^2 d\eta^2-\gamma_{\alpha\beta}(dx^\alpha+N^\alpha d\eta)(dx^\beta+N^\beta d\eta)
\end{equation}
where $N$ and $N^\alpha$ denote  the lapse function and the shift-vector, respectively, while $\gamma_{\alpha\beta}$ ($\alpha,\beta=1,2,3$) is the 3-metric tensor of the spatial hyper-surfaces $\Sigma^3$ $\eta=const$, and we take \cite{K93}
\begin{equation}
\label{parametrizzazione della metrica}
	\gamma_{\alpha\beta}=e^{q^a}\delta_{ad}O^a_b O^d_c \displaystyle\frac{\partial y^b}{\partial x^\alpha} \displaystyle\frac{\partial y^c}{\partial x^\beta},\ \ \ 
	a,b,c,d,\alpha,\beta=1,2,3,
\end{equation} 
where $q^a=q^a(x,\eta)$ and $y^b=y^b(x,\eta)$ are six scalar functions, and $O^a_b=O^a_b(x)$ a $SO(3)$ matrix (repeated indexes are summed).\\
Using the metric tensor (\ref{parametrizzazione della metrica}), the action for the gravitational field reads 
\begin{equation}
\label{azione standard}
	S=\int_{\Sigma^{(3)}\times\Re}d\eta d^3 x\left(p_a\partial_\eta q^a+\Pi_d\partial_\eta y^d -NH-N^\alpha H_\alpha\right)\,,
\end{equation}
with
\begin{equation}	
	\label{vincoli Hamiltoniani} 
	H=\frac{1}{ \sqrt \gamma}\left[\sum_a (p_a)^2-\frac{1}{2}\left(\sum_b p_b\right)^2-\gamma ^{(3)}R\right]
\end{equation}	
\begin{equation}
\label{vincoli Hamiltoniani2}	 
	 H_\alpha=\Pi_c \partial_\alpha y^c +p_a \partial_\alpha q^a +2p_a(O^{-1})^b_a\partial_\alpha O^a_b;
\end{equation}
$H$ and $H_\alpha$ being the super-Hamiltonian and the super-momentum, respectively.\\
In (\ref{vincoli Hamiltoniani}) and (\ref{vincoli Hamiltoniani2}) $p_a$ and $\Pi_c$ play the role of conjugate momenta to the variables $q^a$ and $y^b$, respectively, and $^{(3)}R$ is the Ricci 3-scalar which behaves like a potential term.\\
No sooner should we adopt $y^a$ like new spatial variables, than the Super-momentum constraint $H_\alpha=0$ (provided by the variation of (\ref{azione standard}) with respect to $N^\alpha$)  can be solved, and the action rewrites as\cite{BM04}
\begin{equation}
	\label{finale non approssimata}
	S=\int_{\Sigma^{(3)}\times\Re}d\eta d^3 y \left(p_a\partial_\eta q^a+2p_a(O^{-1})^c_a\partial_\eta O^a_c-NH\right)\,.
\end{equation} 
Furthermore the potential term appearing in the Super-Hamiltonian (\ref{vincoli Hamiltoniani}) can be approximated toward the singularity as
\begin{equation}
\label{pot}
	U=\sum_a{\Theta(Q_a)}\, ,
\end{equation}
where
\begin{equation}\label{funzioni generalizzate}
	\Theta(x)=\begin{cases}
	+\infty\ &  $if$\  x>0,\cr 0\ &  $if$\  x<0,\cr
	\end{cases}
\end{equation}
\begin{equation}
Q_a=\displaystyle\frac{q^a}{\sum_a q^a}\,.
\end{equation}
This picture arises from the vanishing of the metric tensor determinant  close to the singularity, and the quantities $Q_a$'s, known in the literature as {\it anisotropy parameters}, cut a closed domain $\Gamma_Q$ to which the dynamics is restricted.\\
By virtue of the potential structure (\ref{pot}), the Universe dynamics evolves independently in each space point inducing a corresponding factorization for the phase space of the model.\\
Since in $\Gamma_Q$ the potential $U$ asymptotically vanishes, close to the singularity the relation $\partial p_a/ \partial\eta=0$ holds and then the term $2p_a(O^{-1})^c_a\partial_\eta O^a_c$ in (\ref{finale non approssimata}) behaves like an exact time-derivative and can be ruled out of the variational principle.\\ 
The ADM reduction is completed by introducing the so-called Misner-Chitr\'e-like variables \cite{K93,IM01,C72,M00},whose relevance  consists in making  the anisotropy parameters independent of $\tau$, which will behave as the time variable: 
\begin{equation}
	\label{cambio di chitre}
	\begin{cases}
	q^1=e^\tau \left[\sqrt{\xi^2-1}(\cos\theta+\sqrt 3 \sin\theta)-\xi\right]\cr
	q^2=e^\tau \left[\sqrt{\xi^2-1}(\cos\theta-\sqrt 3 \sin\theta)-\xi\right]\cr
	q^3=-e^\tau\left(\xi+2\sqrt{\xi^2-1}\cos\theta\right)\,.
	\end{cases} 
\end{equation}
In terms of these new variables the super-Hamiltonian rewrites
\begin{equation}
\label{Hamiltoniana ridotta}
H=-p_\tau^2+p_\xi^2 (\xi^2-1)+\frac{p_\theta^2}{\xi^2-1}
\end{equation}
and the $Q_a$'s become
\begin{eqnarray}
Q_1=& \displaystyle\frac{1}{ 3}-\frac{\sqrt{\xi^{2}-1}}{ 3\xi}(\cos\theta+\sqrt{3}\sin\theta)\nonumber\\
\label{parametri di anisotropia32}
Q_2=& \displaystyle\frac{1}{ 3}-\frac{\sqrt{\xi^{2}-1}}{ 3\xi}(\cos\theta-\sqrt{3}\sin\theta)\\
Q_3=& \displaystyle\frac{1}{ 3}+\frac{2\sqrt{\xi^{2}-1}}{ 3\xi}\cos\theta \phantom{(\cos\theta+41)}\nonumber
\end{eqnarray}
Let's solve the constraint $H=0$ (obtained variating $N$ in (\ref{azione standard})) with respect to $p_\tau$ in the domain $\Gamma_Q$ in order to perform the ADM reduction, thus obtaining:
\begin{equation}
	\label{Hamiltoniana ADM}
	-p_\tau\equiv\epsilon=\sqrt{(\xi^2-1)p_\xi^2+\frac{p_\theta^2}{\xi^2-1}}\,.
\end{equation}
Then, taking the time gauge $\partial_\eta\tau=1$, the reduced action explicitly reads as
\begin{equation}
\label{azione ridotta}
S_{\Gamma_Q}=\int d\tau d^3 y \left(p_\xi\partial_\tau\xi+p_\theta\partial_\tau\theta-\epsilon\right)\,.
\end{equation}
As we approach the singularity, $\epsilon$ behaves like a constant of motion, i.e. $d\epsilon/ d\tau=\partial\epsilon/\partial\tau=0\Rightarrow \epsilon=E(y^a)$.\\
The dynamics of such a model is equivalent to (the one of) a billiard-ball on a Lobatchevsky plane\cite{CB83, GRAHAM95, BM04, B82}; this can be shown by the use of the Jacobi metric, {\it i.e.} a geometric approach that reduces the dynamical equations of a generic system to a geodesic problem on a manifold. Such a technique  applied to (\ref{azione ridotta})  produces a geodesic equation corresponding to the line element
\begin{equation}
\label{metrica di jacobi}
dl^2=E^{2}(y^a)\left(\frac{d\xi^{2}}{ (\xi^{2}-1)}+(\xi^{2}-1)d\theta^{2}\right)\,
\end{equation}
The manifold described by (\ref{metrica di jacobi}) turns out to have a constant negative curvature, where the Ricci scalar is given by $R=-2/E^2$: the complex dynamics of the generic inhomogeneous model results in a collection of decoupled dynamical systems, one for each point of the space, and all of them equivalent to a billiard problem on a Lobatchevsky plane.\\
Among the possible representations for it, we choose  the so-called Poincar\'e model in the complex upper half-plane \cite{KM97p} that can be introduced with the following coordinate transformation
\begin{equation}
\label{poincare variable}
\begin{cases}
\xi=\displaystyle\frac{1+u+u^2+v^2}{\sqrt{3} v}\\
\theta=-\arctan\displaystyle\frac{\sqrt{3} (1+2 u)}{-1+2 u+ 2u^2+2 v^2}
\end{cases}
\end{equation}
The line element for this 2-dimensional surface now reads
\begin{equation}
	\label{elemento di linea u e v}
	ds^2=\frac{du^2+dv^2}{v^2}
\end{equation}
Figure (\ref{pot1.eps}) shows how, in terms of these coordinates, the anisotropy functions cut a stripe in the plane $(u,\;0,\;v)$ characterized by a finite measure $\mu=\pi$ and three unstable directions to asymptotically escape.\\ The boundary is given by the set of points where one of the anisotropy parameters is equal to zero; it is a geodesic triangle, where the edges are given by the vanishing of the numerators of $Q_a$'s, now reading
\begin{eqnarray}
	\label{anisotropie geodetiche}
	\phantom{i}&\left\{ \begin{array}{ll}
	Q_1(u,v)=-u/\delta\\
	Q_2(u,v)=(1+u)/\delta\\
	Q_3(u,v)=(u^2+u+v^2)/\delta\\
	\end{array} \right.\\
	&\delta=u^2+u+1+v^2\nonumber
\end{eqnarray}
In the $(u,v)$ scheme, the ADM Hamiltonian (\ref{Hamiltoniana ADM}) assumes the expression
\begin{equation}
\label{Hamiltoniana uv}
H_{ADM}=\epsilon=v \sqrt{p_u^2+p_v^2}\,,
\end{equation}
which will be the starting point of our analysis.\\
We conclude by observing how \cite{BKL70,BKL82} the present representation of the Lobatchevsky plane is suitable to link this Hamiltonian formulation to the piece-wise representation provided by Belinski, Khalatnikov and Lifshitz  map, as it is easy to recognize that for $v=0$, the functions $Q_a$ reduce to the familiar Kasner indexes.

\section{Hamilton-Jacobi Approach}

We shall now derive the Hamilton-Jacobi (HJ) \cite{Ar89} equations for the system (\ref{Hamiltoniana uv}) in view of the following developments.\\
The HJ prescribes  to change the momentum variables into the derivatives of a functional ${\mathcal S}$ with respect to the configurational variable: implementing the HJ technique, the Hamiltonian relation (\ref{Hamiltoniana uv}) leads to the functional differential equation in $\Gamma_Q$
\begin{equation}
\label{prescrizione HJ}
-\frac{\delta {\mathcal S}}{\delta \tau}=v\sqrt{\left({\frac{\delta {\mathcal S}}{\delta u}}\right)^2+\left({\frac{\delta {\mathcal S}}{\delta v}}\right)^2}\,.
\end{equation}
To obtain the solution of this dynamical equation, we take ${\mathcal S}$ in the form
\begin{equation}
	\label{funzione S}
	{\mathcal S}={\mathcal S}_0(u,v)-\left(\int d^3 y\epsilon(y^a)\right)\tau ,
\end{equation}
where ${\mathcal S}_0$ satisfies the equation
\begin{equation}
\label{equazione per S}
\epsilon^2 = v^2 \left(\left({\frac{\delta {\mathcal S}_0}{\delta u}}\right)^2+\left({\frac{\delta {\mathcal S}_0}{\delta v}}\right)^2\right)
\end{equation}
and it is therefore provided by
\begin{equation}
\label{S}
{\mathcal S}_0(u,v)=k(y^a) u + \sqrt{\epsilon^2-k^2(y^a) v^2} -\epsilon \ln \left(2\frac{\epsilon+\sqrt{\epsilon^2-k^2(y^a) v^2}}{\epsilon^2 v}\right)+c(y^a)\,,
\end{equation}
where $k(y^a),c(y^a)$ are integration constants.\\
The expression (\ref{funzione S}), together with (\ref{S}) and the features of the potential wall (\ref{anisotropie geodetiche}), summarizes the classical dynamics of a generic inhomogeneous Universe.

\section{The Statistical Mechanic Description}

In the previous Section 2 we have pointed out that, for the Mixmaster inhomogeneous dynamics, the spatial points decouple approaching the singularity and an energy-like constant of motion appears; let us discuss the problem from a statistical mechanics point of view  by treating the system as a microcanonical ensemble. \\
The physical properties of a stationary ensemble are described by a distribution function $\rho=\rho(u,v,p_u,p_v)$, obeying the continuity equation defined in the phase space $(u,v,p_u,p_v)$  as 
\begin{equation}
\label{eq di continuita def}
\displaystyle\frac{\partial (\dot{u} \rho)}{\partial u}+\displaystyle\frac{\partial (\dot{v} \rho)}{\partial v}+\displaystyle\frac{\partial (\dot{p_u} \rho)}{\partial p_u}+\displaystyle\frac{\partial (\dot{p_v} \rho)}{\partial p_v}=0\,,
\end{equation}
where the dot denotes the time derivative
\begin{eqnarray}
\dot{u}\equiv\displaystyle\frac{\partial u}{\partial \tau}=\displaystyle\frac{\partial H_{ADM}}{\partial p_u}=\displaystyle\frac{v^2}{\epsilon} p_u&\hspace{2.0cm}\dot{p}_u\equiv\displaystyle\frac{\partial p_u}{\partial \tau}=-\displaystyle\frac{\partial H_{ADM}}{\partial u}=0\nonumber\\
\label{equazioni di hamilton}
\dot{v}\equiv\displaystyle\frac{\partial v}{\partial \tau}=\displaystyle\frac{\partial H_{ADM}}{\partial p_v}=\displaystyle\frac{v^2}{\epsilon} p_v&\hspace{2.5cm} \dot{p}_v\equiv\displaystyle\frac{\partial p_v}{\partial \tau}=-\displaystyle\frac{\partial H_{ADM}}{\partial v}=-\displaystyle\frac{\epsilon}{v}\,.
\end{eqnarray}
We stress how the above continuity equation provides an appropriate representation for the ensemble associated to the Mixmaster only when we are sufficiently close to the initial singularity and therefore the infinite-potential wall approximation works; in fact, such a model for the potential term corresponds to deal with the energy-like constant of the motion and fixes the microcanonical nature of the ensemble. From a dynamical point of view this picture arises naturally because the Universe volume element vanishes monotonically (for non stationary correction to this scheme in the Misner-Chitr\'e like variables see \cite{M01}).\\
We are interested in studying the distribution function in the $(u,v)$ space, and thus we will reduce the dependence on the momentum variables by integrating $\rho(u,v,p_u,p_v)$ in the momenta space and using the informations contained in the HJ approach. If we assume $\rho$ to be a regular, vanishing at the infinity of the phase space, limited function of its arguments,  and, if use the (\ref{S}), we work out the following equation for $\tilde{w}(u,v;k)=\int\rho(u,v,p_u,p_v) dp_u dp_v$ 
\begin{equation}
\label{eq w}
\displaystyle\frac{\partial \tilde{w}(u,v;k)}{\partial u}+\sqrt{\left(\frac{E}{k v}\right)^2-1}\displaystyle\frac{\partial \tilde{w}(u,v;k)}{\partial v}+\frac{E^2-2 k^2v^2}{k v^2} \frac{\tilde{w}(u,v;k)}{\sqrt{E^2-(k v)^2}}=0
\end{equation}
admitting a solution in terms of a generic function $g$ of the form
\begin{equation}
	\label{soluzione eq w}
	\tilde{w}(u,v)=\frac{g\left(u+v\sqrt{\frac{E^2}{k^2 v^2}-1}\right)}{v \sqrt{E^2-k^2 v^2}}
\end{equation}
The distribution function in $(u,v)$ is obtained after the integration over the constant $k$. 
Indeed, this constant expresses the freedom of choosing the initial conditions (\ref{equazioni di hamilton}), which cannot affect the chaotic properties of the model. Therefore we define the reduced distribution $w(u,v)$ as
\begin{equation}
w(u,v)\equiv\int_A \tilde{w}(u,v;k)dk\,,
\end{equation}
where the integration is over the classical available domain for $p_u\equiv k$
\begin{equation}
\label{dominio k}
	A\equiv \left[-\frac{E}{v},\frac{E}{v}\right]\,.
\end{equation}
It is easy to verify that the microcanonical Liouville measure \cite{BKL70,KM97p,CB83,IM01,BM04,M01,IM03} $w_{mc}$ (after integration over the admissible values of $\epsilon$ ) corresponds to the case $g=const$, {\it i.e.} we get the normalized distribution
\begin{equation}
	\label{misura microcanonica}
	w_{mc}(u,v)=\int_{-\frac{E}{v}}^{\frac{E}{v}}\frac{1}{k v^2 \sqrt{\frac{E^2}{k^2 v^2}-1}}dk=\frac{\pi}{v^2}\,.
\end{equation}
Summarizing, we have derived the generic expression of the distribution function for our model, fixing its form for the microcanonical ensemble which, in view of the energy-like constant of the motion $\epsilon$, is the most appropriate to describe the Mixmaster system when the picture is restricted to the configuration space.

\section{Quasi-classical Limit of the Quantum Regime}

Let us underline some common features between the classical and the semi-classical dynamics with the aim of fixing the proper operator-ordering \cite{KU81} in treating the quantum approach.\\
In fact, considering the WKB limit for $\hbar \to 0$, the coincidence between  the classical distribution function $w_{mc}(u,v)$ and the quasi-classical probability function $r(u,v)$ takes place for a precise choice of the operator-ordering \cite{IM05} only (for a connected topic see \cite{KM97J} and for a general discussion see \cite{Gutz}).\\
As soon as we implement the canonical variables into operators, as in the canonical quantization, {\it i.e}
\begin{equation}
\nonumber
\hat{v}\rightarrow v,\,\,\,\hat u\rightarrow u
\end{equation}
\begin{equation}
\nonumber
\hat{p}_v\rightarrow-i\hbar\displaystyle\frac{\partial}{\partial v},\,\,\,\,\,\,\,\,\hat{p}_u\rightarrow-i\hbar\displaystyle\frac{\partial}{\partial u},\,\,\,\,\,\,\,\, \hat{p}_\tau\rightarrow-i\hbar\displaystyle\frac{\partial}{\partial \tau}\,,
\end{equation}
the quantum dynamics for the state function $\Phi(u,v,\tau)$ obeys, in each point of the space, the Scrh\"odinger equation\\
\begin{equation}
\label{scroedinger}
i\hbar \frac{\partial \Phi(u,v,\tau)}{\partial \tau}=\hat{H}_{ADM}\Phi(u,v,\tau) =\hbar\sqrt{-v^2\displaystyle\frac{\partial^2}{\partial u^2}-v^{2-a}\displaystyle\frac{\partial}{\partial v}\left(v^a \displaystyle\frac{\partial}{\partial v}\right)}\Phi(u,v,\tau)\,,
\end{equation}
where we have adopted a generic operator-ordering for the position and momentum operators parametrized by the constant $a$ \cite{KT90}.\\
In the above equation the Hamiltonian operator has a  non-local character as a consequence of the square root function; in principle this  constitutes a subtle question about the quantum implementation though. We will make the well-grounded ansatz\cite{Puzio} that the operators $\hat{H}_{ADM}$ and $\hat{H}_{ADM}^2$ have the same set of eigenfunctions with eigenvalues $E$ and $E^2$, respectively\footnote{The problems discussed in this respect by \cite{KMV} do not arise here because in the domain $\Gamma_Q$ our ADM Hamiltonian has a positive sign (the potential vanishes asymptotically)}.\\

If we take the following  integral representation for the wave function $\Phi$
\begin{equation}
	\label{trasf fourier}
	\Phi(u,v,\tau)=\int_{-\infty}^{\infty} \Psi(u,v,E)e^{-{\mathit i}E\tau/\hbar}d E\,,
\end{equation}
the eigenvalues problem reduces to
\begin{equation}
\label{hquadro}
\hat H^2 \Psi(u,v,E)= \hbar^2\left[-v^2\displaystyle\frac{\partial^2}{\partial u^2}-v^{2-a}\displaystyle\frac{\partial}{\partial v}\left(v^a \displaystyle\frac{\partial}{\partial v}\right)\right]\Psi(u,v,E) =E^2 \Psi(u,v,E)\,.
\end{equation}
In order to study the WKB limit of equation (\ref{hquadro}), we separate the wave function into its phase and modulus
\begin{equation}
\label{forma funzionale psi}
\Psi(u,v,E)=\sqrt{r(u,v,E)} e^{i \sigma(u,v,E)/\hbar}\,.
\end{equation}
In this scheme the function $r(u,v)$ represents the probability density, and the quasi-classical regime appears as we take the limit for $\hbar\to 0$; substituting (\ref{forma funzionale psi}) in (\ref{hquadro}) and retaining only the lower order in $\hbar$, we obtain the system
\begin{equation}
\label{hj system}
\begin{cases}
v^2 \left[\left(\displaystyle\frac{\partial \sigma}{\partial u}\right)^2+\left(\displaystyle\frac{\partial \sigma}{\partial v}\right)^2\right]=E^2\cr
\displaystyle\frac{\partial r}{\partial u} \displaystyle\frac{\partial \sigma}{\partial u} +\displaystyle\frac{\partial r}{\partial v} \displaystyle\frac{\partial \sigma}{\partial v}+r\left(\displaystyle\frac{a}{v} \displaystyle\frac{\partial \sigma}{\partial v} +\displaystyle\frac{\partial^2 \sigma}{\partial v^2} +\displaystyle\frac{\partial^2 \sigma}{\partial u^2}\right)=0\,.
\end{cases}
\end{equation}
In view of the HJ equation and of the Hamiltonian (\ref{Hamiltoniana uv}), we can identify the phase $\sigma$ to the functional $S_0$ of the HJ approach.\\
Now we turn our attention to the equation for $r(u,v)$; taking (\ref{S}) into account it reduces to
\begin{equation}
\label{eq per r}
k \displaystyle\frac{\partial r(u,v,E)}{\partial u}+\sqrt{\left(\frac{E}{v}\right)^2-k^2} \displaystyle\frac{\partial r(u,v,E)}{\partial v}+\frac{a(E^2-k^2 v^2)-E^2}{v^2 \sqrt{E^2-k^2 v^2}} r(u,v,E)=0\,.
\end{equation}
Comparing (\ref{eq per r}) with (\ref{eq w}), we easily see that they coincide (as expected) for  $a=2$ only.\\
It is worth noting that this correspondence is expectable once a suitable  choice for the configurational variables is taken; however here it is remarkable that it arises only if the above operator-ordering is addressed. It is just in this result the importance of this correspondence whose request fixes a particular quantum dynamics for the system.\\
Summarizing, we have demonstrated from our study that it is possible to get a WKB correspondence between the quasi-classical regime and the ensemble dynamics in the configuration space, and we provided the operator-ordering when quantizing the inhomogeneous Mixmaster model to be \cite{KT90}
\begin{equation}
	\label{ordinamento operatoriale}
	\hat{v}^2 \hat{p}_v^2\,\rightarrow\,-\hbar^2 \frac{\partial}{\partial v}\left( v^2\frac{\partial}{\partial v}\right)\,.
\end{equation}
By means of these results, we now face the full quantum dynamical approach.\\

\section{Canonical Quantization and the Energy Spectrum}

\begin{figure}[h]
\begin{center} 
\includegraphics[width=6.0cm]{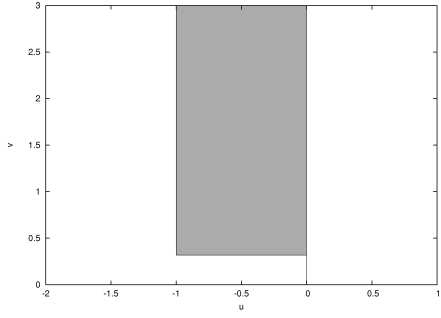} 
\caption{The approximate domain where we impose the boundary conditions.\protect\\
The choice $v=1/\pi$ for the straight line preserves the measure $\mu=\pi$\label{pot approx}} 
\end{center}
\end{figure}

Our starting point is the point-like eigenvalue equation (\ref{hquadro}) (for a discussion of  this same eigenvalue problem in the Misner variables see \cite{M69}) which, together with the boundary conditions, completely describes the quantum features of the model, {\it i.e.}
\begin{eqnarray}
\label{autoval}
	&\left[v^2 \displaystyle\frac{\partial^2}{\partial u^2}  +v^2 \displaystyle\frac{\partial^2}{\partial v^2} +2 v \displaystyle\frac{\partial}{\partial v}+\left(\displaystyle\frac{E}{\hbar}\right)^2\right] \Psi(u,v,E)=0\cr
 &\Psi (\partial \Gamma_Q)=0\,.
\end{eqnarray}
In equation (\ref{autoval}) we can recognize a well known operator: by redefining $\Psi(u,v,E)=\psi(u,v,E)/v$, we can reduce (\ref{autoval}) to the eigenvalue problem for the Laplace-Beltrami operator in the Poincar\'e plane
\begin{equation}
	\label{lb equation}
	\nabla_{LB} \psi(u,v,E)\equiv v^2 \left(\displaystyle\frac{\partial^2}{\partial u^2}+\displaystyle\frac{\partial^2}{\partial v^2} \right) \psi(u,v,E)=E_s\psi(u,v,E)\,,
\end{equation}
which is central in the harmonic analysis on symmetric spaces and has been widely investigated in terms of its invariance under $SL(2,C)$(for a detailed discussion, see \cite{Terras} and the bibliography therein).\\
Its eigenstates and eigenvalues are described as
\begin{eqnarray}
	\label{forma generica delle autofunzioni}
 &\psi_s(u,v)= a v^s+ b v^{1-s}+  \sqrt{v}\sum_{n\neq0}a_n K_{s-1/2}(2\pi |n| v) e^{2\pi {\mathit i} n u},\,\,\,a,b,a_n \in C \cr
 \label{condizioni al bordo}
  &\nabla_{LB} \psi_s(u,v)=s(s-1) \psi_s(u,v) 
\end{eqnarray}
where  $K_{s-1/2}(2\pi n v)$ are the modified Bessel functions of the third kind, and $s$ denotes the index of the eigenfunction.
This is a continuous spectrum, and the summation runs over every real value of $n$.\\
The eigenfunctions for our model read as
\begin{equation}
	\label{autofunzione bianchi}
	\Psi(u,v,E)=a v^{s-1}+ b v^{-s} +\sum_{n\neq 0} a_n \frac{K_{s-1/2}(2\pi|n|v)}{\sqrt{v}} e^{2 \pi{\mathit i} n u}
\end{equation}
with eigenvalue
\begin{equation}
	\label{autovalore bianchi}
	E^2=s(1-s)\hbar^2\,.
\end{equation}
The spectrum and the explicit eigenfunctions are obtained by imposing the boundary conditions (\ref{autoval}), {\it i.e.} requiring that equation (\ref{autofunzione bianchi}) vanishes on the edges of the geodesic triangle of Fig.\ref{pot1.eps}.\\
Since from our analysis no way arose to impose exact boundary conditions we approximate the domain with the simpler one in Fig.\ref{pot approx}; the value for the horizontal line $y=1/\pi$ provides the same measure for the exact and the approximate domain.\\
The conditions on the vertical lines $u=0$, $u=-1$ require to disregard the first two terms in (\ref{autofunzione bianchi}) ($a=b=0$); furthermore we get the condition on the last term\\ 
\begin{equation}
\nonumber
\sum_{n\neq 0} e^{2\pi {\mathit i} n u} \to \sum^{\infty}_{n=1} \sin(2\pi n u),
\end{equation}
  $n$ being an integer, while the condition on the horizontal line implies
\begin{equation}
	\label{relazione sulle k}
	\sum_{n>0}a_n K_{s-1/2}(2n) \sin (2n\pi u)=0, \forall u \in [-1,0]\,,
\end{equation}
which is satisfied by requiring $K_{s-1/2}(2n)=0$ only.\\
The functions $K_\nu(x)$ are real and positive for real argument and real index, therefore the index must be imaginary, {\it i.e}  $s=\displaystyle\frac{1}{2}+i t$. In this case these functions have (only) real zeros, and the corresponding eigenvalues turn out to be real and positive 
\begin{figure}[ht]
 \begin{minipage}[b]{\textwidth}
	\begin{center}
	\includegraphics[width=0.6\textwidth]{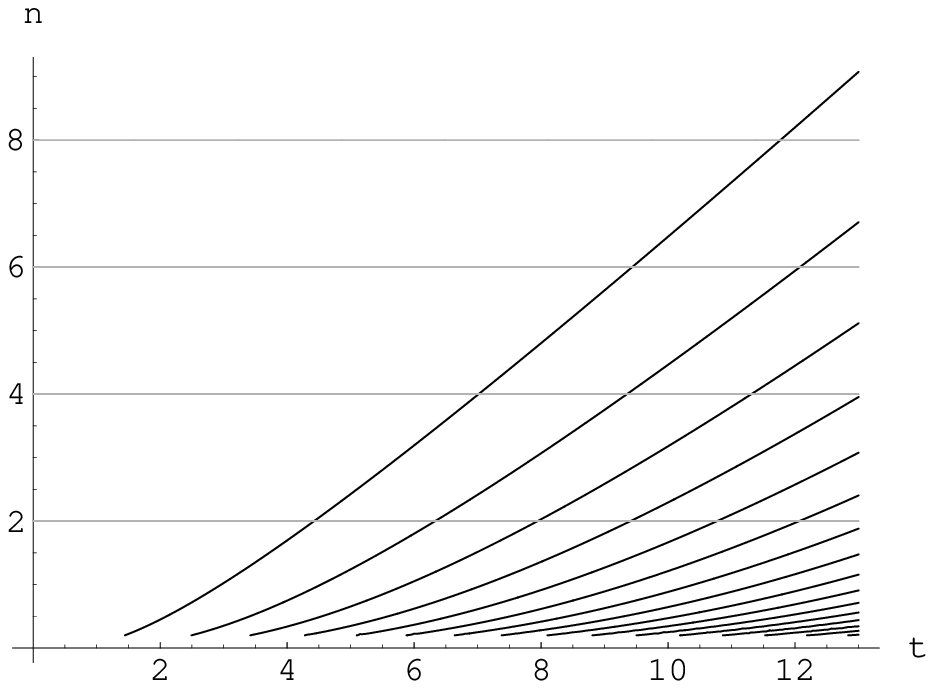}
      \makeatletter\def\@captype{figure}\makeatother 
      \caption{The intersections between the straight lines and the curves represents the roots of the equation $K_{i t}(n)=0$, where $K$ is the modified Bessel function \label{zeri}}
	\end{center}
	\begin{center}
      \begin{tabular}[t]{|c|c|c|c|}
				\hline 
				\textbf{n=2}  &\textbf{n=4}  &\textbf{n=6}  &\textbf{n=8}\\
				\hline
				\hline 
				4.425   &7.016 & 9.434&11.768\\
    		6.333   & 9.353& 12.076& 14.655\\
    		7.947   & 11.313& 14.283 &17.059\\
				9.410  & 16.264& 16.263&19.212\\
				10.774 & 17.742& 18.096&21.203\\
				\hline
		 \end{tabular}
  \end{center}
		\makeatletter\def\@captype{table}\makeatother
		\caption{We numerically investigated the roots of $K_{i t}(n)=0$ with respect $t$ for different values of $n$. The values in the columns are the first $t$'s in the above figure.}
    \label{tabella zeri}
    \end{minipage}
\end{figure}
\begin{figure}[ht]
   \begin{minipage}[b]{0.5\textwidth}
			 \centering
      \begin{tabular}{|c|}
					\hline
					\textbf{\mbox{$\left(\displaystyle\frac{E}{\hbar}\right)^2=t^2+\displaystyle\frac{1}{4}$}} \\
					\hline
					\hline 
   				19.831\\
   				40.357\\
   				49.474\\
   				63.405\\
					87.729\\
					89.250\\
					116.329\\
					128.234\\
					138.739\\
					146.080\\
					\hline
				\end{tabular}
				\vspace{0.7cm}
				\makeatletter\def\@captype{table}\makeatother 
      \caption{The first ten energy eigen-\protect\\values, ordered from the\protect\\ lowest one.}
			\label{primi livelli}
			\end{minipage}%
	\begin{minipage}[b]{0.5\textwidth}
     \centering
     \includegraphics[width=0.6\textwidth,clip]{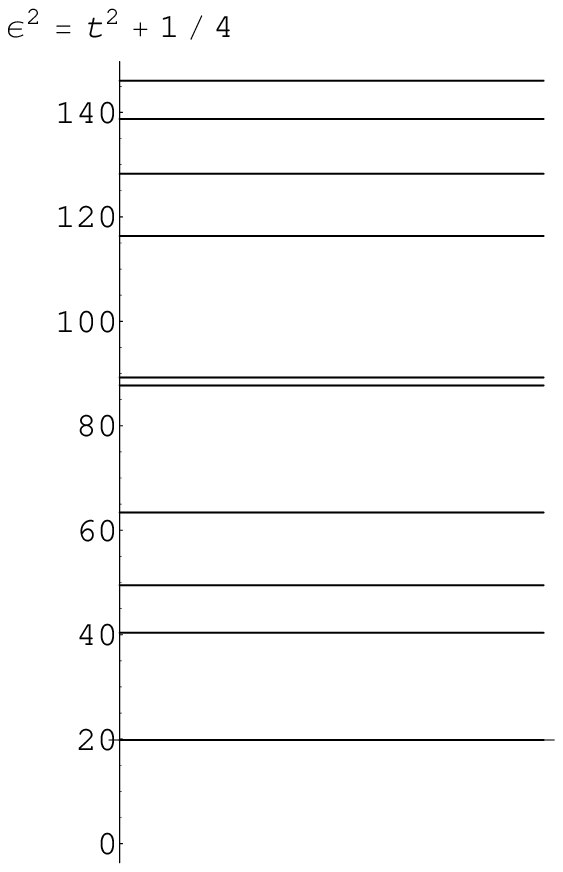}
      \makeatletter\def\@captype{figure}\makeatother 
      \caption{The lines represent in a typical spectral manner the first levels.}
      \label{linee}
      \vspace{0.4cm}	
   \end{minipage}
\end{figure}

\begin{equation}
	\label{spettro finale}
	(E/\hbar)^2=t^2+1/4\,.
\end{equation}
We remark that our eigenfunctions naturally vanish as infinite values of $v$ are approached.\\
The conditions (\ref{relazione sulle k}) cannot be solved analytically for all the values of $n$ and $t$, and the roots must be worked out numerically for each $n$; there are several results on their distribution that allow us to find at least the first levels.
A theorem on the zeros of these functions states that $K_{{\mathit i} \nu}(\nu x)=0$ $\Leftrightarrow$ $0<x<1$ (for a proof see \cite{Palm}); by this theorem and the monotonic dependence of the energy (\ref{spettro finale}) on the zeros, we can easily search the lowest levels by solving equation (\ref{relazione sulle k}) for the first $n$. In Fig.\ref{zeri} and Tab \ref{tabella zeri} we plot the first roots, and in Fig.\ref{linee} and in Tab.\ref{primi livelli} we list the first ten "energy" levels\footnote{
for a detailed numerical investigation of the energy spectrum of the standard Laplace-Beltrami operator, especially with respect to the high-energy levels,   see\cite{CGS91a, CGS91d}, where it is also numerically analyzed the effects on the level spacing of deforming the circular boundary condition toward the straight line}.\\
Below we will provide an analytical treatment of the high energy levels associated to our operator in correspondence to certain region of the plain $(t,\;0,\;n)$. In \cite{GRAHAM95} the structure of the high energy levels is also connected to the so-called \emph{quantum chaos} of the Mixmaster model, which was studied from the wave functional point of view in \cite{FU86} and from the the path integral one in \cite{BE89} using Misner variables. Our analysis implicitly contains  the information about the quantum chaos of the considered dynamics; in fact we can take a generic state of the system $\xi(\tau,\;u,\;v)$ in the form
\begin{equation}
\label{sviluppo}
	\xi(\tau,\;u,\;v)=\int dt\sum_{n} c_{t,n} \Psi_{n}(\tau,\;u,\;v)\delta(K_{i t}(2 n))e^{-i \sqrt{\frac{1}{4}+t^2} \tau}
\end{equation}
and its evolution from a generic initial condition $\xi_0(u,\;v)\equiv \xi(\tau_0,\;u,\;v)$ at an initial instant $\tau_0$ provides all the quantum properties of the system. The quantum chaos is recognized in  \cite{FU86} by numerically integrating the Wheeler- De Witt equation from a gaussian like initial packet and outlining the appearance of a fractal structure in the profile of the resulting wave function; in our approach (as the infinite potential walls approximation works) the dynamics is provided by (\ref{sviluppo}) and evolving it from an initial localized wave packet the quantum chaos has to arise.\\
About the information on the quantum chaos emerging from the high-energy spectrum we emphasize the following two important points: i) the non-stationary corrections due to the real potential term are expected to be simply small perturbation to our result rapidly decaying as the singularity is approached; ii) the analytic expressions we are going to provide for large values of $t$ cannot be used to fix the existence of the quantum chaos because they explore limited regions of the plain $(t,\;0,\;n)$ only, but they are very useful to clarify the morphology of the spectrum and its dependence on two different quantum numbers.

\subsection{The Ground State}

\begin{figure}[ht]
	\begin{minipage}[b]{0.5\textwidth}
		\centering 
			\includegraphics[width=1.1\textwidth,clip]{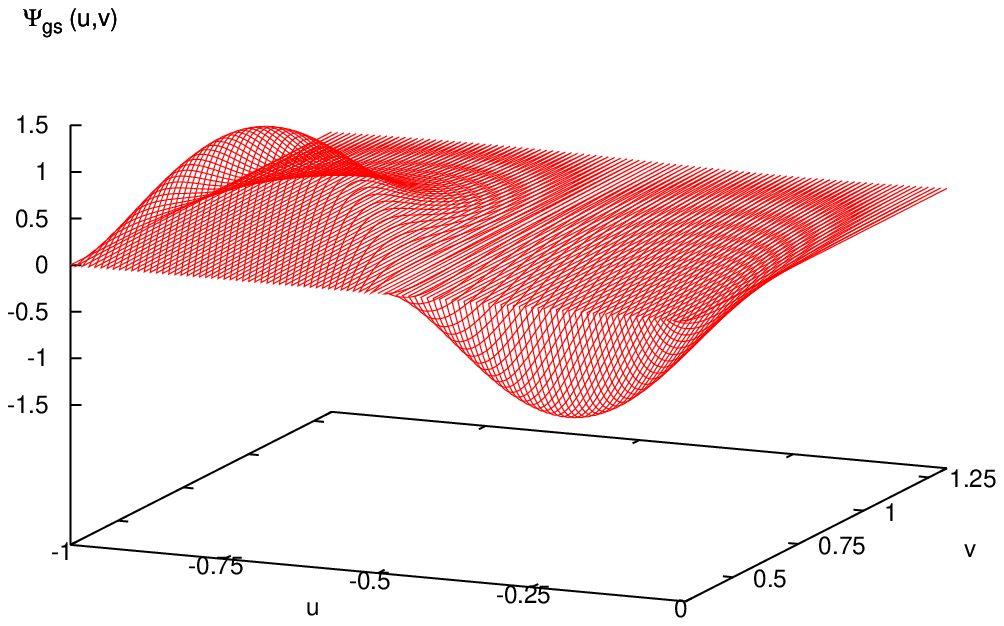} 
			\caption{The ground state wave function\protect\\ is sketched \label{groundstate} }
			\vspace{0.95cm}
	\end{minipage}%
	\begin{minipage}[b]{0.5\textwidth}
			\centering 
			\includegraphics[width=1.1\textwidth,clip]{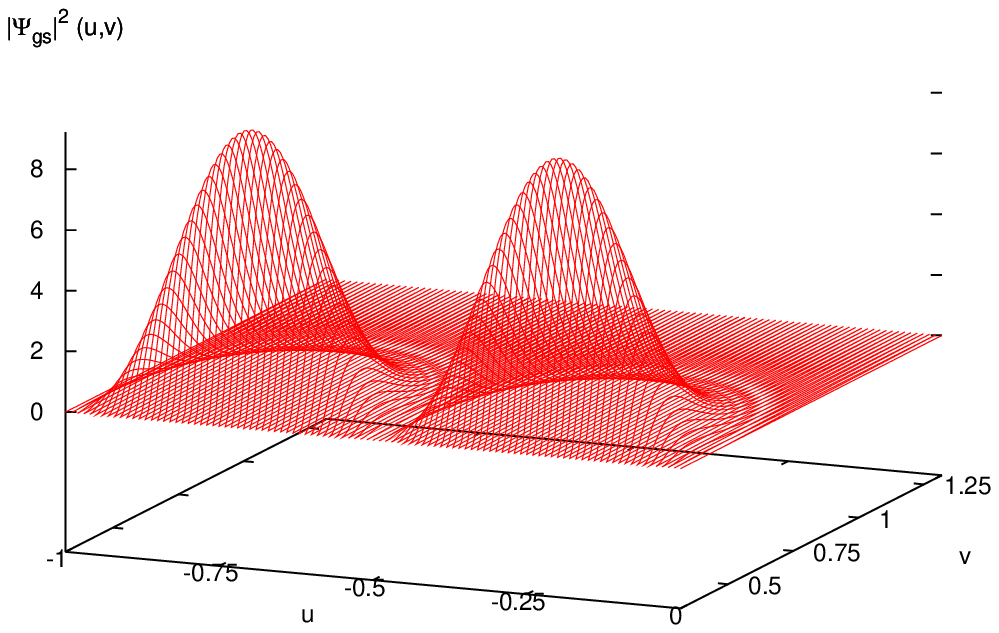}
			\caption{We show the probability distribution in correspondence to the ground state of the theory, according to the boundary conditions. \label{probabilita}} 
	\end{minipage}
\end{figure}

Let us describe the properties of the ground state level with a major accuracy, starting from the result of its existence with a non-zero energy $E_0$, {\it i.e.} $E_0^2=19.831 \hbar^2$. \\
In Fig.\ref{groundstate} we plot the wave function $\Psi_{gs}$ in the $(u,\;0,\;v)$ plane and in Fig.\ref{probabilita} the corresponding probability distribution $|\Psi_{gs}|^2$, whose normalization constant is equal to $N=739.466$.
The knowledge of the ground state eigenfunction allows us to estimate the average values of the anisotropy variables $u$, $v$ and the corresponding root mean square, {\it i.e.} 
\begin{equation}
	\langle u \rangle=-1/2\;,\hspace{1.0cm}\langle v \rangle=0.497
\end{equation}
\begin{equation}
\Delta u=0.266\;,\hspace{1.0cm}\Delta v=0.077
\end{equation}
This result tells us that in the ground state the Universe is not exactly an isotropic one, and it fluctuates around the line of the Misner plane $\beta_-=0$. However we observe that it remains localized in the center of the Misner space far from the corner at $v\to\infty$ (the other two equivalent corners were cut out from our domain by the approximation we considered on the boundary conditions, but, because of the potential symmetry, they have to be unaccessible too). Thus we can conclude that the Universe, approaching the minimal energy configuration, conserves a certain degree of anisotropy but lives in the region where the latter can be treated as a small correction to the full isotropy.
Such a behavior is a consequence of the zero-point energy associated to the ground state which prevents the absence of oscillation modes concerning the anisotropy degrees of freedom; we numerically derived this feature but it can be inferred on the basis of general considerations about the Hamiltonian structure. In fact the Hamiltonian contains a term $\hat{v^2p^2_v}$ which has positive definite spectrum and cannot admit vanishing eigenvalue.

\subsection{Asymptotic Expansions}

In order to study the distribution of the highest energy levels, let us take into account the asymptotic behavior of the zeros for the modified Bessel functions of the third kind.\\
We will discuss asymptotic regions of the plane $(t,\;0,\;n)$ in both the cases i) $t\gg n$ and \\ii) $t\simeq n\gg1$.\\
\vspace{0.4cm}\\
i) For $t\gg n$ the Bessel functions admit the following representation:
\begin{equation}
\label{sviluppo asintotico tutto a destra}
K_{i t}(n)=\frac{\sqrt{2\pi}e^{-t\pi/2}}{(t^2-n^2)^{1/4}}\left[ \sin a\sum_{k=0}^\infty \frac{(-1)^k}{t^{2k}} u_{2k}(\frac{1}{\sqrt{1-p^2}})+\cos a \sum_{k=0}^\infty \frac{(-1)^k}{t^{2k+1}} u_{2k+1}(\frac{1}{\sqrt{1-p^2}})\right]
\end{equation}
where $a=\pi/4-\sqrt{t^2-n^2}+t\,$arccosh$(t/n)$, $p\equiv n/t$ and $u_k$ are the following polynomials
\begin{equation}
	\left\{ \begin{array}{ll}
	u_0(t)=1\\
	u_{k+1}(t)=\displaystyle\frac{1}{2} t^2(1-t^2) u_k'(t)+\frac{1}{8}\int_0^1(1-5t^2)u_k(t)dt\\
	\end{array} \right.
\end{equation}
Retaining in the above expression only those terms of $\mathsf{o} (\frac{n}{t})$, the zeros are fixed by the following relation
\begin{equation}
	\label{arcotangente}
	\sin(\frac{\pi}{4}-t+t (\log(2)-\log(p)))-\frac{1}{12 t} \cos(\frac{\pi}{4}-t+t (\log(2)-\log(p)))=0
\end{equation}
that in the limit for $n/t\ll 1$, can be recast as follows
\begin{equation}
	\label{arcotangente2}
	t \log(t/n)=l \pi\;\Rightarrow\; 	t= \frac{l \pi}{\mathrm{productlog}(\frac{l \pi}{n})}
\end{equation}
where productlog(z) is a generalized function that gives the solution of the equation $z=w e^w$ and for real and positive domain, it is a monotonic function of its argument. In (\ref{arcotangente2}) $l$ is an integer number that must be taken much greater than 1 in order to verify the initial assumptions $n/t\ll 1$.\\
ii) In case the difference between $2n$ and $t$ is $\mathsf{o}(n^{1/3}) (t,n\gg 1$), we can evaluate the first zeros by the expansion (worked out from formula (9) of \cite{Balogh})
\begin{equation}
  \label{zeri asintotici}
  t= 2n + 0.030 n^{1/3}\,,
\end{equation}
providing the lowest zero (and therefore the energy) for a fixed  value of $n$ and then the relation for the eigenvalues for high occupation numbers as
\begin{equation}
\label{alti numeri}
 \left(\frac{E}{\hbar}\right)^2\sim 4 n^2 + 0.12 n^{4/3} 
\end{equation}
In this region of the spectrum the condition $t\sim 2n$, reduces the 2 quantum numbers characterizing the system to a single one, the whole wave function becoming assigned by $n$.\\
The above analysis shows that, as effect of Dirichlet boundary condition and in the limit of high occupation numbers, we get an analytical expressions for the discrete structure of the spectrum. In fact  for large values  of $t$ it was possible to give analytical representations  for the position of the zeros, but we emphasize that the request to deal with these approximations causes the loss of a large number of levels and prevents a complete discussion of the quantum chaos associated to the model. However the above expressions are of interest because allow to compare these results  with the corresponding spectrum provided by Misner in his original work \cite{M69q}. Of course, such a comparison of the two results can take place only on a qualitative level; in fact, between Misner ($\alpha,\; \beta_+,\; \beta_-$) and Misner-Chitr\'e like ($\tau,\; u,\; v$)  variables a crucial difference exists and it has to be recognized into the correspondingly different behavior of the potential walls. In the Misner scheme  the domain available to the point Universe increases as $\alpha\to\infty$ ($\alpha$ being the isotropic Misner variable) and, therefore, we deal with a non stationary infinite well; the Misner-Chitr\'e like variables allow to fix the infinite potential walls into a time-independent configuration in the $(u,\,0,\,v)$ plane. However we can at least compare the behavior of the energy eigenvalues with respect to the occupation number $n$.\\
In his original treatment, Misner replaces, for fixed $\alpha$ values (indeed he makes use of an adiabatic approximation, see \cite{IM05}), the triangular well by a square of equal measure and determines the energy spectrum in the form
\begin{equation}
	\frac{E_n^M}{\hbar}=\frac{\pi}{\sqrt{S}}\sqrt{n_+^2+n_-^2}\equiv \frac{A}{\alpha}|n|,
\end{equation}
where $S$ denotes the triangular well measure $S=\alpha^2/A^2$, $A$ being a numerical factor; above $n_\pm$ denote  the occupation numbers relative to $\beta_\pm$ respectively.\\
In our approach, for sufficiently large $n$, we get the dominant behaviors 
\begin{equation}
	\frac{E_n^{MC}}{\hbar}\sim \begin{cases}
	$i)$\frac{l \pi}{\mathrm{productlog}(\frac{l \pi}{n})}\\
  $ii)$2 n + \mathcal O(n^{2/3})
	\end{cases}
\end{equation}
Thus we see that, in case ii), a part from a numerical factor  (which is different because of the different approximation made on the real domains), the only significant difference relies on the term $\alpha^{-1}$; in the case i) instead the situation is a bit different because 2 quantum numbers explicitly remain and we get a linear relation as far as we require $l\propto n$ by a factor much greater than the unity (indeed the function productlog$(l\pi/n)$ provides a smooth contributions in the considered region $l\gg n$).
The difference of the factor $\alpha^{-1}$ with respect to the Misner case can be easily accounted as soon as we observe that the following relations hold:
\begin{equation}
\label{gio}
	\beta_\pm= \alpha b_\pm (u,\;v)\;,
\end{equation}
where the functions $b_\pm$ can be calculated from the coordinates transformations 
\begin{equation}
	\left\{\begin{array}{l}
	\alpha=-e^\tau \displaystyle\frac{1+u+u^2+v^2}{\sqrt{3} v}\\
	\beta_+=e^\tau \displaystyle\frac{-1+2u+2u^2+2v^2}{2\sqrt{3} v}\\
	\beta_-=-e^\tau \displaystyle\frac{1+2u}{2 v}
	\end{array}\right.
\end{equation}
 but their form is  not relevant in what follows. On the base  of (\ref{gio}), the measure of a domain $D$ in the $\beta_\pm$ plane reads
\begin{equation}
	\int_D d\beta_+d\beta_-=\alpha^2 \int_{D'}|J(u,v)| du dv\;,
\end{equation}
$J$ being the Jacobian of the transformation associated to $b_\pm$, while $D'$ is the image of $D$ onto the $(u,\;0,\;v)$ plane. As a consequence we see that between a measure $s$ in the $\beta_\pm$ plane and a similar one (even not exactly its image), there is a difference for  a factor $\alpha^2$ which immediately provides an explanation for the difference in the 2 energy spectra.

\section{The Inhomogeneous Picture and Conclusions}

At the end of our analysis, we wish to bring the reader's attention to some physical aspects of the inhomogeneous Mixmaster.\\
First of all, the obtained dynamics regime is indeed a generic one; in fact, in a synchronous reference (for which $\partial_t y^a=0$), the inhomogeneous Mixmaster contains four independent (physically) arbitrary functions of the spatial coordinates ($y^a(x^i)$ and $E(x^i)$), available for the Cauchy data on a non-singular hyper-surface.\\
Let us now come back to the full inhomogeneous problem, in order to understand the structure of the quantum space-time near the cosmological singularity. Since the spatial gradients of the configurational variables play no relevant dynamical role in the asymptotic limit $\tau\to\infty$ (indeed the spatial curvature simply behaves as a potential well) then the quantum evolution above outlined takes place independently in each space point and the total wave function of the Universe can be represented as follows
\begin{equation}
\label{funzione d'onda dell'universo}
	\Xi (\tau,\;u,\;v)= \Pi_{x_i} \xi_{x_i} (\tau,\;u,\;v)
\end{equation}
where the product is (heuristically) taken over all the points of the spatial hypersurface.\\
However, it is worth to recall that, in the spirit of the "long-wavelength approximation" adopted here, the physical meaning of a space point must be recovered on the notion of a cosmological horizon; in fact we are dealing with regions over which the inhomogeneity effects are negligible and this statement corresponds to super-horizon sized spatial gradients. On a classical point of view, this request is at the ground of the BKL approximation and it is well confirmed on the statistical level (see \cite{K93}); however on a quantum level it can acquire a precise meaning  if we refer the dynamics to a kind of lattice space-time in which the spatial gradients of the configurational variables become real potential terms. In this respect it is important to observe that the geometry of the space-time is expected to acquire a discrete structure on the Planck scale  and we believe that a regularization of our approach could arrive from a "loop quantum gravity" treatment \cite{Bojowald:2003md}.\\
Despite this local homogeneous framework of investigation, the appearance near the singularity of high spatial gradients and of a space-time foam (like outlined in the classical dynamics see \cite{K93,M95}) can be easily recognized in the above quantum picture too.
In fact the probability that in $n$ different space points (horizons) the variables $u$ and  $v$ take values within the same narrow interval, decrease with $n$ as $p^n$, $p$ being the probability in a single point; in fact, these probabilities are all identical to each other and no interference phenomenon takes place. From a physical point of view, this simple consideration indicates that a smooth picture of the large scale Universe is forbidden on a probabilistic level and different causal regions are expected completely disconnected from each other during their quantum evolution.
Therefore, if we start with a strongly correlated initial wave function $\Xi_0(u,\;v)\equiv\Xi(\tau_0,\;u,\;v)$, its evolution toward the singularity induces increasingly irregular distributions, approaching (\ref{funzione d'onda dell'universo}) in the asymptotic limit $\tau\to\infty$.\\
The main result of our presentation can be recognized in the clear correspondence established between the classical space-time foam and the quantum one. We have outlined how this link takes place naturally for a precise choice of the operator-ordering only and how the "energy spectrum" is a discrete one, due to the billiard structure of the point-like Hamiltonian.
Finally, we fixed as a new feature the zero-point "energy"  for the ground state associated to the anisotropy degrees.

\section*{Acknowledgments}

We'd like to thank Orchidea Maria Lecian and Gian Paolo Imponente for revising the manuscript.

\end{document}